\documentclass[aps,prl,twocolumn,superscriptaddress]{revtex4}
\usepackage{graphicx}
\usepackage{color}
\usepackage{amssymb}
\usepackage{amsmath}
\usepackage{siunitx}
\usepackage{ulem}
\usepackage{textcomp}
\usepackage{booktabs}

\newcommand{\ket}[1]{|#1\rangle}

\definecolor{mygreen}{rgb}{0,0.5,0} 
\definecolor{myred}{rgb}{0.75,0,0} 
\definecolor{myblue}{rgb}{0,0,0.75} 
\definecolor{mymagenta}{cmyk}{0,1,0,0.12} 
\definecolor{mycyan}{cmyk}{1,0,0,0.12} 
\definecolor{myorange}{rgb}{1,0.5,0}

\newcommand{\btext}[1]{{\color{myblue}#1}}

\newcommand{\gtext}[1]{{\color{mygreen}#1}}
\renewcommand{\gtext}[1]{{\color{mygreen}}}

\newcommand{\ntext}[1]{{\color{black}#1}}

\newcommand{\op}[1]{\hat{#1}}
\newcommand{\agg}[1]{\hat{#1}^{\dag}}
\newcommand{\media}[1]{\langle #1 \rangle}
\newcommand{\E}[2]{\op{a}_{ #1} (#2)}
\newcommand{\Ea}[2]{\agg{a}_{ #1} (#2)}
\newcommand{\R}[3]{R_{#1,#2}^{ (2)} (#3)}
\newcommand{\Rs}[2]{R_{#1,#2}^{ (2)} }

\newcommand{\nC}{{\Phi_\textsf{C}}}
\newcommand{\nS}{{\Phi_\textsf{S}}}
\newcommand{\DMtwo}{\rho^{(2)}}

\newcommand{\commacite}[1]{ \cite{#1},}
\newcommand{\periodcite}[1]{ \cite{#1}.}
\newcommand{\inlinecite}[1]{ \cite{#1}}

\usepackage{siunitx}

\begin{document}

\title{

A macroscopic quantum state analysed particle by particle\\

}

\author{Federica A. Beduini}
\affiliation{ICFO -- Institut de Ciencies Fotoniques, Mediterranean Technology Park, 08860 Castelldefels (Barcelona), Spain}

\author{Joanna A. Zieli\'{n}ska}
\affiliation{ICFO -- Institut de Ciencies Fotoniques, Mediterranean Technology Park, 08860 Castelldefels (Barcelona), Spain}

\author{Vito G. Lucivero}
\affiliation{ICFO -- Institut de Ciencies Fotoniques, Mediterranean Technology Park, 08860 Castelldefels (Barcelona), Spain}

\author{Yannick A. de Icaza Astiz}
\affiliation{ICFO -- Institut de Ciencies Fotoniques, Mediterranean Technology Park, 08860 Castelldefels (Barcelona), Spain}

\author{Morgan W. Mitchell}
\affiliation{ICFO -- Institut de Ciencies Fotoniques, Mediterranean Technology Park, 08860 Castelldefels (Barcelona), Spain}
\affiliation{ICREA -- Instituci\'{o} Catalana de Recerca i Estudis Avan\c{c}ats, 08015 Barcelona, Spain}

\date{26 October 2014}

\begin{abstract}
We report particle-by-particle measurements on a macroscopic quantum state. We analyse the joint polarisation state of photon pairs extracted randomly from a beam of polarisation-squeezed light, an archetypal macroscopic quantum system analogous to squeezed states in spin-1/2 atomic ensembles.  We confirm several predictions of recent spin-squeezing theory [Beduini {\it et al.} Phys. Rev. Lett. 111, 143601 (2013)] including  entanglement  among all pairs of photons arriving within the squeezing coherence time, ``NooN''-type entanglement, and concurrence that decreases with photon flux, as required by entanglement monogamy.  
\end{abstract}

\maketitle



   Explaining how microscopic entities collectively produce
macroscopic phenomena is a fundamental goal of
many-body physics. Theory predicts that large-scale
entanglement is responsible for exotic macroscopic 
phenomena\commacite{BardeenPR1957, AndersonS1987,SorensenN2001, SorensenPRL2001} 
but  observation of entangled
particles in naturally occurring systems is extremely
challenging.  Synthetic quantum systems made of atoms in optical
lattices\inlinecite{BlochN2008} have been constructed with the goal of observing
macroscopic quantum phenomena with single-atom 
resolution\periodcite{BakrN2009,ShersonN2010} 
Serious challenges remain in producing  
and detecting 
 long-range quantum correlations in these systems, however\periodcite{GeorgescuRMP2014} Here we exploit the strengths of 
photonic technology, including high coherence and efficient single-particle detection, 
to study the predicted large-scale entanglement\inlinecite{BeduiniPRL2013} underlying the macroscopic quantum phenomenon 
of polarization squeezing\periodcite{KorolkovaChapter2007} 
  We generate a polarization-squeezed beam, extract photon pairs at random, and make a tomographic reconstruction\inlinecite{AdamsonPRL2007} of their joint quantum state.  We present experimental evidence showing that all photons arriving within the squeezing coherence time are entangled, that entanglement monogamy dilutes entanglement with increasing photon density and that, counter-intuitively, increased squeezing can reduce bipartite entanglement.  The results provide direct evidence for entanglement of macroscopic numbers of particles\inlinecite{BeduiniPRL2013} and introduce micro-analysis to the study of macroscopic quantum phenomena.

Squeezing phenomena, in which fluctuations in a macroscopic variable are reduced below the na\"{i}ve quantum limits, offer a privileged window into the role of entanglement in macroscopic quantum phenomena.  Spin-squeezing inequalities (SSIs) demarcate the boundary between squeezed states and classical states, i.e. between macroscopic behaviour producible with and without entanglement.  Beyond detecting entanglement\commacite{GuhnePR2009} SSIs can quantify entanglement depth\inlinecite{SorensenPRL2001}
and the number of entangled particles\periodcite{VitaglianoPRL2011}  Spin squeezing experiments\inlinecite{GrossN2010,BehboodPRL2014} have used SSIs to claim 500,000 entangled atoms and entanglement depth of 170.  These claims far exceed the records with individually-controlled particles: 14 trapped ions\inlinecite{MonzPRL2011} and 8 photons\periodcite{YaoNPhot2012}

Here we report the first study of entangled particles underlying a macroscopic quantum phenomenon, and the  first test of an SSI as a predictor of microscopic entanglement.  While SSIs have  mostly been applied to atomic spin squeezing\commacite{GrossN2010,BehboodPRL2014} a direct test with atoms appears challenging.  In contrast to individually trapped atoms/ions, detection of atoms from ensembles\inlinecite{NelsonNPhys2007, BuckerNJP2009, BakrN2009, ShersonN2010} has not yet shown the simultaneous, individual, state-selective  detection required for multi-particle state characterisation\periodcite{JamesPRA2001}

Photons are becoming an attractive system for studying strongly correlated bosons, showing Bose-Einstein condensation (BEC)\commacite{KlaersN2010} nonlinear Josephson oscillations\commacite{AbbarchiNPhys2013}  dynamical squeezing\commacite{MikhailovOL2008} and strong, Rydberg-atom-mediated  interactions\periodcite{FirstenbergN2013} Recent work has probed photon-BEC thermodynamics through photon-number statistics\inlinecite{SchmittPRL2014} and there are proposals to use photons for lattice-gas quantum simulators\periodcite{KlaersSPIE2013}   {Here we add the capability for micro-state analysis to the photonic quantum gas toolkit.} We use the photon polarisation as a binary degree of freedom, analogous to spin-1/2 atoms.

\begin{figure*}
\includegraphics[width=0.9\textwidth]{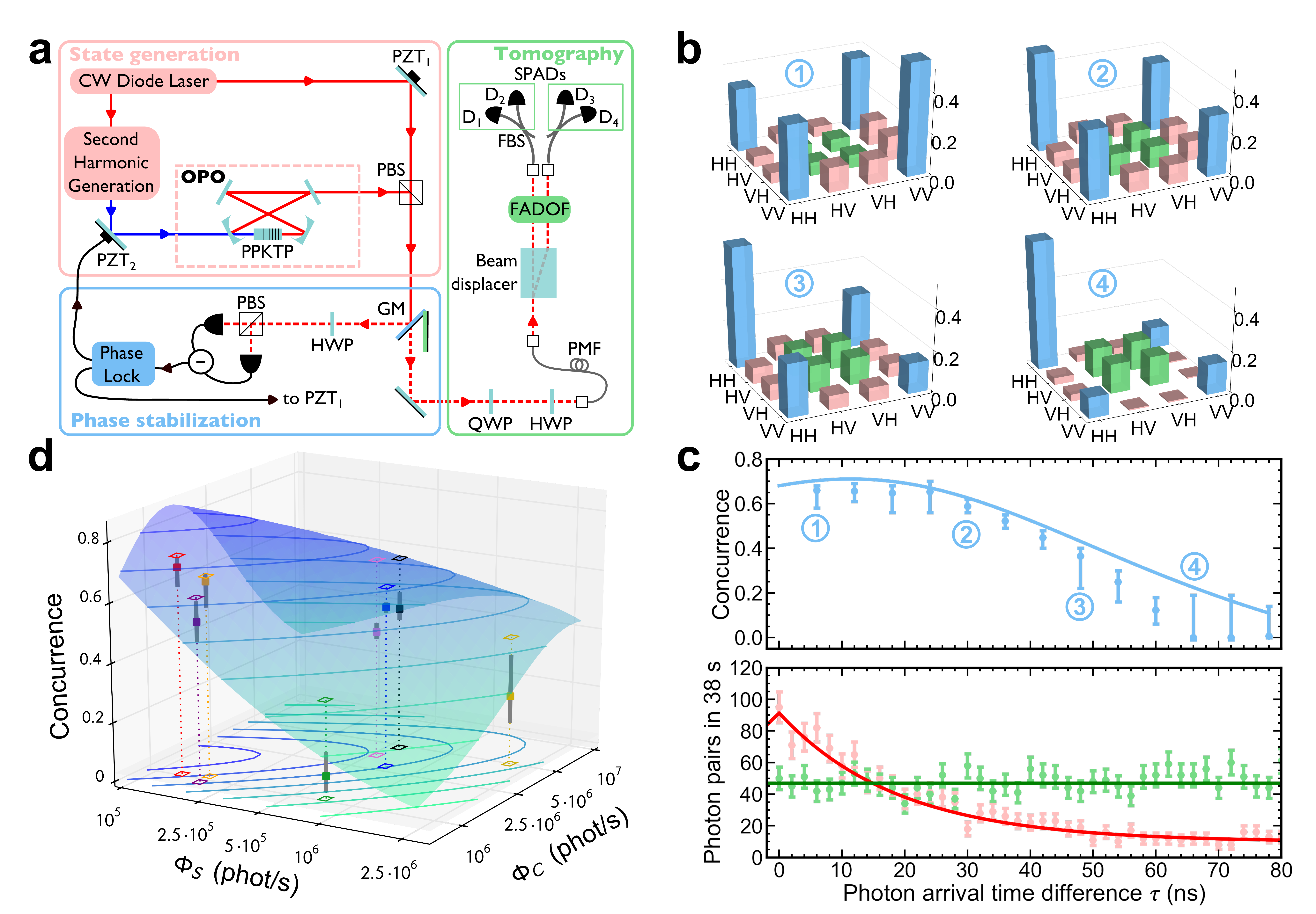}

\caption{
{
Photon-level analysis of a polarisation-squeezed state.
 {\bf a}:  Experimental schematic.  (red box) A 795~nm continuous-wave laser is locked to the Rb D$_1$ line, frequency doubled, and used to pump a sub-threshold optical parametric oscillator (OPO), producing 795~nm squeezed vacuum (SV) with vertical (V) polarisation.  A portion of the laser is also taken to produce a horizontal (H) polarisation coherent state (CS), and the two components are combined on a polarising beamsplitter (PBS) to produce polarisation-squeezed (PS) light.  Two piezo-electric actuators (PZT)s act on the relative phase of the SV and CS components of the state.  (blue box) A galvanometer mirror (GM) directs the PS beam alternately to a continuous-variable (CV) detector, sensitive to the macroscopic polarisation of the state, or to a discrete-variable (DV) photon-counting system.  The CV detector consists of a half-wave plate (HWP), PBS, and linear differential detector.  Its electronic signals are fed back to PZT$_1$ and PZT$_2$ to stabilise the SV/CS relative phase.  (green box) the DV detector consists of a quarter-wave plate (QWP), HWP, polarisation-maintaining fibre and beam displacer, together acting as a polarisation analyser splitting the two polarisations (in a basis determined by the QWP and HWP) to two parallel output channels.  A Faraday anomalous-dispersion optical filter (FADOF) is used to eliminate photons from non-degenerate modes of the OPO.  The two output paths are each split to two single-photon avalanche photodiodes (SPADs) to allow detection of both same- and opposite-polarisation photon pairs.  {\bf b}:
Reconstructed density matrices {(magnitudes only)} of photons extracted from a PS state with $\nC = 9.6 \times 10^5$~ph/s and $\nS = 1.9 \times 10^5$~ph/s {(orange square in Fig.~\ref{fig:big}{\bf d})} for mean arrival time differences {$\langle{|\tau|}\rangle
 = 6, 30, 48$ and $66$~ns.} We include all events {within the coincidence window $\langle{|\tau|}\rangle - 6 \btext{~\text{ns}} \leq {|\tau|} \leq \langle{|\tau|}\rangle + 6$~ns. } Blue bars indicate the ``NooN'' portion of the state, a superposition of $\ket{\textsf{HH}}$ and $\ket{\textsf{VV}}$, green bars indicate the ``W'' portion of the state, $\propto \ket{\textsf{HV}}+\ket{\textsf{VH}}$, and pink bars indicate anomalous coherences (see text).  {\bf c}:  (upper) Concurrence of the two photon density matrices versus {$\langle{|\tau|}\rangle $} under the same conditions.  Error bars indicate $\pm 1 \sigma$ statistical error estimated by a bootstrapping procedure (see Methods).  Results show entanglement for photons separated by up to 60~ns, as predicted by the theory (solid line).  (lower) SV (red) and CS (green) contributions to the PS state.  Shown are photon pair detection rates as a function of delay time for one or the other contribution, indicating a cross-over in source brightness at $\tau = 15$~ns, which corresponds to balanced $\ket{\textsf{HH}}$, $\ket{\textsf{VV}}$ amplitudes and maximum concurrence. {\bf d}: Comparison between the theoretical concurrence (surface) and the experimental observations (filled squares) for a coincidence time window of 26~ns centred on ${\tau}=0$. The contour plot and the empty squares on the bottom plane are the projection of the theoretical surface and of the experimental data on the space of SV and CS photon fluxes, $\nS$ and $\nC$, respectively. The upper empty squares lie on the surface and represent the expected concurrence for the measured density matrices. Grey bars indicate $\pm 1 \sigma$ statistical errors calculated by bootstrapping. For all cases, we obtain theoretical predictions by integrating the elements of the expected density matrix $\rho$ over the time window considered.
}
}
\label{fig:big}
\end{figure*}

~

\ntext{
{\noindent Theory --}
Non-classical polarisation correlations and photon entanglement 
are related through the second-order  correlation functions $\R{ij}{mn}{\tau}\equiv\media{\Ea{i}{t}\Ea{j}{t+\tau}\E{n}{t+\tau}\E{m}{t}}$, where the $\Ea{i}{t}$ are mode operators and subscripts indicate H or V polarisation\periodcite{BeduiniPRL2013}  Classically, these obey the Cauchy-Schwarz-like inequalities 
\begin{subequations}
	\begin{eqnarray}
		|\R{\text{\text{HH}}}{\text{VV}}{\tau}|^2 &\le& \R{\text{HV}}{\text{HV}}{\tau}\R{\text{VH}}{\text{VH}}{\tau}
		\label{eq:G2HHVVa}   \\
		|\R{\text{HV}}{\text{VH}}{\tau}|^2 &\le& \R{\text{\text{HH}}}{\text{HH}}{\tau} \R{\text{VV}}{\text{VV}}{\tau},
		\label{eq:G2HVHVb}
	\end{eqnarray}
\end{subequations}
which can be violated by squeezed fields.  At the same time, these correlation functions give the  two-photon density matrix $\DMtwo$ via Glauber photodetection theory, 
$\DMtwo_{ij,kl}(\tau) \propto \R{ij}{kl}{\tau}$\periodcite{BeduiniPRL2013}  For a {polarisation squeezed (PS)} state consisting of {vertically}-polarised {(V)} squeezed vacuum and a stationary {horizontally}-polarised {(H) coherent} field, the density matrix is a so-called X-state  
in the $\{\text{HH,HV,VH,VV}\}$ basis, i.e. with non-null elements only along the two diagonals:
\begin{equation}
   \DMtwo \propto
   \begin{pmatrix}
      \Rs{\text{HH}}{\text{HH}} &   0   &   0   &   \Rs{\text{HH}}{\text{VV}} \\
      0 &   \Rs{\text{HV}}{\text{HV}} &   \Rs{\text{HV}}{\text{\text{VH}}}  &   0\\
      0 &    \Rs{\text{\text{VH}}}{\text{HV}} &   \Rs{\text{\text{VH}}}{\text{\text{VH}}}  &   0\\
      \Rs{\text{VV}}{\text{HH}}   &   0   &   0   &   \Rs{\text{VV}}{\text{VV}} \\
   \end{pmatrix}\;,
   \label{eq:matrix}
\end{equation}
where we have suppressed the $\tau$ dependence for clarity.  $\DMtwo$ is non-positive under partial transpose, and thus entangled by the Peres-Horodecki criterion, if either of Eqs (\ref{eq:G2HHVVa}), (\ref{eq:G2HVHVb}) is violated, i.e. for polarisation squeezed states.  In other words, this type of polarisation squeezing requires an underlying photonic entanglement.

Density matrices $\DMtwo(\tau)$ are given explicitly in\inlinecite{BeduiniPRL2013} as a function of $\nC$ and $\nS$, the photon fluxes in the coherent and squeezed components of the PS state, respectively.  {The predicted $\DMtwo(\tau)$ shows large concurrence,
~up to 100\%, for pure {squeezed vacuum (SV)} with  low squeezing, i.e. $\nS \ll \Gamma$, where $\Gamma$ is the  bandwidth of the squeezed vacuum\periodcite{BeduiniPRL2013}  In these conditions, the concurrence is large 
for a region defined by 
\begin{equation}
\Gamma \nS \approx \nC^2 , \;\;\;  \nS < \Gamma ,  \;\;\; \tau \Gamma < 1.
\label{eq:SweetSpot}
\end{equation}
}
This geometry reflects the fact that the entanglement arises from two-photon interference, which is strongest when the two-photon contributions from the H and V states are similar, {i.e., when $\DMtwo_\text{HH,HH}(\tau)\approx\DMtwo_\text{VV, VV}(\tau)$}.

}

~ 

{\noindent Experiment -- } The apparatus to produce PS states and detect photon pairs extracted from them is shown in Fig.~\ref{fig:big}{\bf a}. The state is generated using a polarising beam splitter to combine a {H}-polarised coherent state into the same spatial mode as a {V}-polarised squeezed vacuum state with the same frequency. The SV is the output of the degenerate mode of a type-I optical parametric oscillator (OPO)\periodcite{PredojevicPRA2008} The relative phase of the H and V inputs is stabilised with the help of a ``seed beam'' injected into the OPO, as described in the Methods. {A stable phase $\varphi$ between the H and V component is necessary to obtain entangled states: if the corner off-diagonal elements, namely $\rho_\text{HH,VV}$ and its hermitian conjugate $\rho_\text{VV,HH}$, vanish, also the concurrence goes to zero.  As $\rho_\text{HH,VV}\propto e^{i\varphi}$, when the phase is fluctuating freely during the photon acquisition, the tomography reconstruction procedure yields the average value of $\rho_\text{HH,VV}$, which is equal to zero in case of fast phase drifts compared to the acquisition time (4/5 hours long). 
During the measurement, we stabilise the length of both the coherent and the OPO pump path with active feedback on the position of one mirror in each path: this maintains $\varphi$ stable within a few degrees for hours.}

{We reconstruct} the polarisation state of the photon pairs belonging to the PS state by discrete quantum tomography\periodcite{JamesPRA2001} 
Our polarisation analyser consists of a quarter-wave plate (QWP) and a half-wave plate (HWP) at angles $\theta_{\rm QWP}, \theta_{\rm HWP}$, respectively, followed by coupling into a polarisation maintaining fibre (PMF) with fast axis aligned to the H polarisation.  The fast and slow polarisations are then separated with a calcite beam displacer, filtered (see below), split with 50/50 fibre beam splitters (FBSs), and detected with single-photon avalanche photodiodes (SPADs).  A multi-channel time-stamping board and post-processing are used to record arrival times. Seven sets of QWP and HWP angles, listed in the Methods, are used to detect in 7 distinct polarisation bases, and for each basis we {can} collect both same-polarisation and mixed-polarisation coincidences.  We  recover the photon pair density matrix with a maximum likelihood algorithm\inlinecite{AdamsonPRL2007} for each  delay $\tau$. 

We use a Faraday Anomalous Dispersion Atomic Filter (FADOF)~\inlinecite{ZielinskaOL2012} to eliminate photons in non-degenerate OPO modes, which would otherwise be detected by the broadband SPADs. The filter has a transmission bandwidth of 223~MHz HWHM, narrower than the OPO's 500~MHz free spectral range, rejects the $\sim 600$ out-of-band modes at the $1:10^{5}$ level, and has been measured to give an output consisting of at least 96\% degenerate-mode photons\periodcite{ZielinskaOE2014}

Both phase stabilisation and quantum tomography involve the measurement of the PS state, but they cannot happen simultaneously, as any attempt at splitting the state would {reduce both the degree of squeezing and its purity}.  We  use a galvanometer mirror to switch rapidly ($\sim$ 100~Hz) the PS state between the tomography and the phase stabilisation setup, to acquire photons in  $\sim$ 3~ms intervals, small compared to the few-second time scale of phase drifts in the system.

~

{\noindent Results -- }%
{Density matrices for different values of $\tau$ are shown in Fig.~\ref{fig:big}{\bf b}, for $\nC = 9.6 \times 10^5$ ph/s and $\nS = 1.9 \times 10^5$ ph/s.}
These show the predicted ``X'' shape, apart from  {small but nonzero coherences off of the two diagonals, e.g.  $\rho_{\text{VV,VH}}.$ These are unexpected, but can be explained as leakage of CS light into the SV polarisation, e.g. by imperfections of the combining PBS or FBS.   The density matrices show strong $\rho_{\text{HH,VV}}$ coherences, giving a good fidelity with a ``NooN''-like state of the form $\cos \theta \ket{\text{HH}} + \sin \theta \ket{\text{VV}}$. Fig.~\ref{fig:big}{\bf c} shows the relationship between concurrence (top graph) and the relative strength of the $\ket{\text{HH}}$ and $ \ket{\text{VV}}$ components (bottom graph). This confirms that  maximum concurrence occurs when {the two-photon interference between H and V polarisation is maximum, i.e. } 
$\rho^{(2)}_\text{HH,HH}(\tau)\approx \rho^{(2)}_\text{VV,VV}(\tau)$ {(see Eq.~\eqref{eq:SweetSpot})}.  Because $\rho^{(2)}_\text{VV, VV}(\tau)$ drops off exponentially with $\tau$ while $\rho^{(2)}_\text{HH,HH}(\tau)$ is $\tau$-independent, the concurrence necessarily drops off for photons widely separated in time.  The concurrence values shown give lower bounds on the entanglement of the photons in the squeezed state: decoherence due to experimental limitations, e.g. noise in the phase stabilisation on the time-scale of the acquisition, would reduce the coherences and thus the entanglement. 
}

{
Fig.~\ref{fig:big}{\bf d} shows measured concurrence  at several values of $\nC$ and $\nS$ with {$|\tau|<13$~ns}, and confirms other aspects of the predicted relationship between squeezing and particulate entanglement.  First, in all cases a statistically significant entanglement is observed.  Second, the concurrence decreases away from the maximum entanglement area described in Eq. (\ref{eq:SweetSpot}), either if the $\nC$ and $\nS$ components are imbalanced, or if $\nS$ increases beyond the squeezing bandwidth.  It is perhaps surprising that more squeezing can give less entanglement.  This is, however, required by entanglement monogamy: each photon is entangled with all of its co-arrivees, but the total entanglement (concurrence) is limited.  Hence the entanglement with any given other photon must decrease. 
}

{Conclusions --}
We report the first particle-by-particle measurements on a macroscopic quantum state. We analyse the joint polarisation state of photon pairs extracted randomly from a beam of polarisation-squeezed light, an archetypal macroscopic quantum system analogous to squeezed states in spin-1/2 atoms.  We confirm several predictions of recent spin-squeezing theory\commacite{BeduiniPRL2013} including strong entanglement, with concurrence up to 0.7,  among all pairs of photons arriving within the squeezing coherence time, ``NooN''-type entanglement, and concurrence that decreases with photon flux as required by entanglement monogamy.  The technique, thus proven, can be applied to particle-by-particle studies of entanglement in interacting and/or computationally-intractable bosonic systems, e.g. photon BECs\commacite{KlaersN2010} exciton polaritons\commacite{AbbarchiNPhys2013} and Rydberg-blockade-bound photon gases\periodcite{FirstenbergN2013}

{ Acknowledgements --}%
We thank Raymond Y. Chiao and Jack Boyce for early discussions, Geoff Lundeen for early exploratory experiments, and David Paredes-Barato for useful discussions. This work was supported by the Spanish MINECO  project MAGO (Ref. FIS2011-23520), by the European Research Council project AQUMET, and by Fundaci\'{o} Privada CELLEX. {J.Z. was supported by the FI-DGR PhD-fellowship program of the Generalitat of Catalonia. Y.A.d.I.A. was supported by the scholarship BES-2009-017461, under the Spanish MINECO Project No.~FIS2007-60179.}

\bibliographystyle{nature}
\bibliography{../biblio/01_Quantum_OpticsBD,../biblio/BigBib130123,../biblio/EEP}


\cleardoublepage
{
\onecolumngrid
   \section*{SUPPLEMENTAL MATERIALS}

\begin{figure*}[t]
\includegraphics[width=0.9\textwidth]{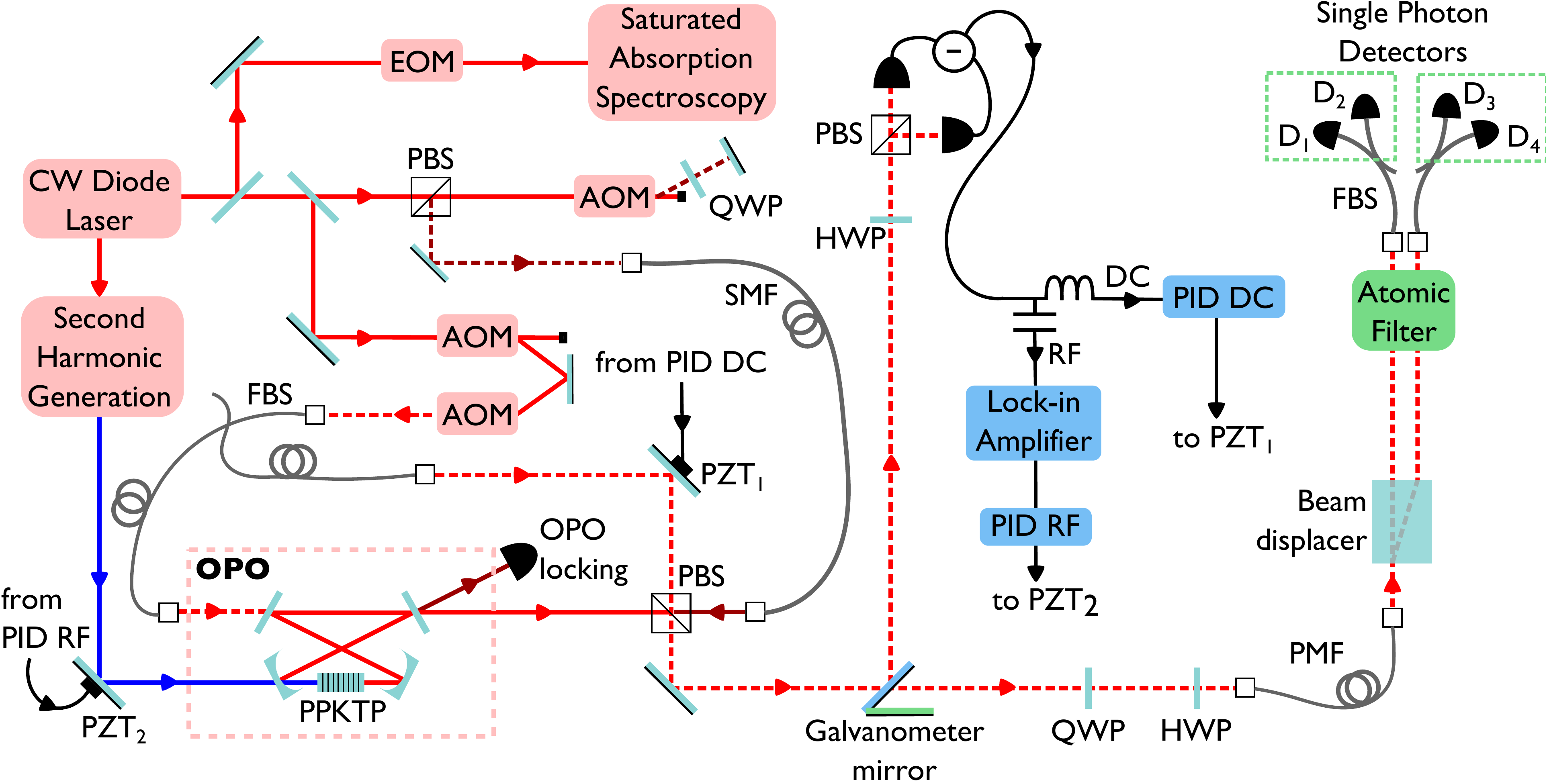}
\caption{Full Setup.  For detailed explanation, see text.  
}
\label{fig:FullSetup}
\end{figure*}

A detailed schematic of the setup is shown in Fig.~\ref{fig:FullSetup}. 

\section*{Quantum light source}

An extended cavity diode laser (ECDL) at 795~nm provides the fundamental optical frequency $\nu_0$ for the experiment, and all optical fields are derived, directly or indirectly, from this source.  We use an integrated electro-optical modulator (EOM) to apply 960 MHz sidebands to a portion of the ECDL output, and lock the lower sideband to the crossover line of the $F=2 \rightarrow F'$ transition of ${}^{85}$Rb using saturated absorption spectroscopy.  This stabilises $\nu_0$ to  2.7~GHz to the red of the centre of the rubidium D${}_1$ line, which is the centre of the transmission window of the FADOF filter.  Light at $\nu_0$ from the ECDL is used as the coherent state (CS) component of the PS, as the ``seed beam'' (see below) and also, after frequency shifting, for stabilisation of the OPO (see below).

{We amplify  and frequency-double} part of the {ECDL} output {to generate a $397$~nm pump for the OPO at $2 \nu_0$.}   {The degenerate mode of the OPO at $\nu_0$ then produces one-mode squeezed vacuum at $\nu_0$ while non-degenerate modes, offset from $\nu_0$ by multiples of the 500 MHz FSR, contain two-mode squeezed vacuum.}  
{
The OPO~\inlinecite{PredojevicPRA2008} is a bow-tie cavity with a 10~mm long potassium titanyl phosphate {(KTP)} crystal, periodically-poled {(PP)} for V-polarised parametric down-conversion. \gtext{and is described in detail in \inlinecite{PredojevicPRA2008}.} We use the Pound-Drever-Hall technique, with feedback to a cavity mirror, to stabilise the cavity length.  To avoid introducing light into the degenerate mode, we perform the stabilisation using a mode that is counter-propagating, H-polarised, and, due to the birefringence of the crystal, shifted in frequency by 630 MHz.  A double-pass AOM setup is used to shift the frequency of the locking beam.  
}

\section*{Phase stabilization}
{We stabilize $\varphi \equiv \varphi_\text{SV} - \varphi_\text{CS}$, the relative phase between the SV and the CS beams, using an intermediate phase reference, a \gtext{weak} V-polarised ``seed beam'' with frequency $\nu_0$, injected into the degenerate mode of the OPO through a cavity mirror.  We note that the phase of the SV derives from $\varphi_{\text{pump}}$, the phase of the OPO pump beam, so {stabilising $\varphi$} \gtext{it} is equivalent to stabilise $\varphi_\text{pump} - 2\varphi_\text{CS}$.  In the OPO, the seed beam experiences phase-sensitive parametric amplification, giving an output with amplitude that depends on $\varphi_\text{pump}- 2\varphi_\text{seed}$, i.e. the relative phase between the pump and the seed beam. When the amplified seed beam is combined with the CS beam at the PBS and detected with a balanced detector in the $\pm 45$\textdegree~polarisation basis, the signal depends also on  $\varphi_\text{seed}-\varphi_\text{CS}$.  To stabilise $\varphi_\text{pump}-2\varphi_\text{seed}$, we modulate $\varphi_\text{pump}$ at $\approx 50$~kHz by moving a mirror mounted on a piezo-electric actuator in the pump path. The balanced detector signal is demodulated with a lock-in amplifier, fed back via a PI controller to the average position of the same mirror.  Similarly, the DC component is fed back to a mirror in the CS beam path, to stabilise $\varphi_\text{seed}-\varphi_\text{CS}$.  With both stabilised, $\varphi$ is constant {within a few degrees}.  }

\section*{Chopped measurement/Stabilization}
{The experiment alternates between periods of data acquisition (DA) and periods of stabilisation (ST) at a rate of {$\approx$}90 Hz. A galvanometer mirror (GM) is used to direct the PS beam toward the polarisation analyser, filter, and SPADs during DA, and toward the continuous-variable polarimeter during ST.  Synchronous with the GM, the AOMs in the seed beam path and in the OPO locking beam path transmit the seed beam and locking beam only during ST.  Also using AOMs, the CS beam is maintained at a few hundred $\mu$W during ST to give a strong signal for phase stabilisation, and attenuated to $10^6$ -- $10^7$ photons/s during DA to study entanglement properties as in Fig. \ref{fig:big}{\bf c}.  Finally, we gate the single photon detectors with a TTL signal so that they are not active during ST.  
We use a measurement duty cycle of 30\%.}

\newcommand{\DD}{P_3}

\section*{Quantum state tomography}

We use a permutationally-invariant state reconstruction\commacite{AdamsonPRL2007} to recover the density matrix from a tomographically-complete set of 10 measurements of photon pair arrival rates for different polarisations.  The polarisation density matrix describing arrival of one photon at time $t$ and another at time $t+\tau$ must be invariant under permutation of the time indices, because the SV and CS contributions to the state are continuous-wave, and because no subsequent optical elements couple arrival time to polarisation.

Using the triplet-singlet basis $\{ HH, \psi^{+}, VV, \psi^{-} \}$, where $\psi^{\pm} \equiv (HV\pm VH)/\sqrt{2}$, we  write a general PI state as $\rho_\text{PI} = L^\dagger L$, where 
\begin{equation}
   L \equiv 
   \begin{pmatrix}
     p_1      & 0   & 0 & 0 \\
     p_5+ip_6 & p_2 & 0 & 0 \\
     p_7+ip_8 & p_9+ip_{10} & p_3 & 0\\
     0 & 0 & 0 & p_4
   \end{pmatrix}\,
\end{equation}
and $p_1\ldots p_{10}$ are real parameters.  {We make 10 independent and tomographically complete measurements described by the positive operator-valued measure (POVM) in the computational basis $\{HH, HV, VH, VV\}$
\begin{equation}
\Pi_m \equiv (U^{(m)}_\text{HWP}U^{(m)}_\text{QWP})^{\otimes 2} P_m (U_\text{QWP}^{\dagger(m)} U_\text{HWP}^{\dagger(m)})^{\otimes 2},
\end{equation} 
where
\begin{eqnarray}
U_\text{HWP} &\equiv& \begin{pmatrix}\cos 2\theta_\text{HWP} & \cos 2\theta_\text{HWP}\\ \sin 2\theta_\text{HWP} & -\cos 2\theta_\text{HWP} \end{pmatrix}, \\
U_\text{QWP} &\equiv& \begin{pmatrix} \cos^2\theta_\text{QWP} + i \sin^2\theta_\text{QWP} & (1-i)\sin\theta_\text{QWP}\cos\theta_\text{QWP}\\ (1-i)\sin\theta_\text{QWP}\cos\theta_\text{QWP} & i\cos^2\theta_\text{QWP} +  \sin^2\theta_\text{QWP} \end{pmatrix}
\end{eqnarray}
are the Jones matrices associated to the quarter- and half-wave plate, respectively, and $P_i$ are projectors associated to the detection of photon pairs with a specific polarisation, i.e. $P_1 = |HH\rangle\langle HH|$, $P_2 = |VV\rangle\langle VV|$ and $P_3 = |HV\rangle\langle HV|+ |VH\rangle\langle VH|$. We compute $\text{Tr}[\Pi_m \rho_{CB}]$, where $\rho_{CB}$ is $\rho_{PI}$ written in the computational basis, to predict the experimental outcomes $n_\text{exp,m}$, i.e. the number of photon pairs $n_\text{exp,m}$~detected during measurement $m$ on the detector pairs shown in the table below, within the specified coincidence time window. 
Thanks to the four SPADs configuration, we can obtain the 10 experimental results $n_\text{exp,m}$ ($m = 1\dots 10$) with just the 7 waveplates settings shown in the list below,  reducing significantly the total acquisition time. 
}

~

\noindent
\hfill
\begin{center}
   \begin{tabular}{ccccc|ccccc}
   \toprule
  m & $\theta_{\rm HWP}$ & $\theta_{\rm QWP}$ & $P$ & Detectors & m & $\theta_{\rm HWP}$ & $\theta_{\rm QWP}$ & $P$ & Detectors\\
   \midrule
    1 & 0        & 0 & $P_{1}$ & D${}_1$D${}_2$ & 6 & $\pi/8$ & 0       & $\DD$ & $\sum_{i=1}^2\sum_{j=2}\text{D}_i\text{D}_j$\\
    2 & 0        & 0 & $P_{2}$ & D${}_3$D${}_4$ & 7 & $\pi/8$ & $\pi/4$ & $\DD$ & $\sum_{i=1}^2\sum_{j=2}\text{D}_i\text{D}_j$ \\
    3 & 0        & 0 & $\DD$  & $\sum_{i=1}^2\sum_{j=2}\text{D}_i\text{D}_j$  & 8 & $\pi/8$ & $\pi/8$ & $P_{1}$ & D${}_1$D${}_2$\\
    4 & $\pi/16$ & 0 & $P_{1}$ & D${}_1$D${}_2$ & 9 & $\pi/4$ & $\pi/8$ & $P_{1}$ & D${}_1$D${}_2$\\
    5 & $\pi/8$  & 0 & $P_{1}$ & D${}_1$D${}_2$ & 10 & 0      & $\pi/8$ & $\DD$ & $\sum_{i=1}^2\sum_{j=2}\text{D}_i\text{D}_j$\\
      \bottomrule
   \label{tab:wp}
\end{tabular} 
   
\end{center}

\hfill

{To correct for unequal efficiencies in the fibre beamsplitters and detectors, we first measure the singles detection rate $\beta_{i,\SI{45}{\degree}}$ arriving to the $i$-th detector  when a 
coherent state with polarisation at 45\textdegree, and thus equal H and V power, is introduced into the PMF.  We compute the normalised path-and-detector efficiency  $\gamma_{i} = \beta_{i,\SI{45}{\degree}}/\sum_j \beta_{j,\SI{45}{\degree}}$. Similarly, we write $\beta_{i,m}$ for the detection rate at the $i$-th detector for setting $m$ during tomography, and define $\alpha_{m} \equiv \sum_i \beta_{i,m}/\gamma_i$, the total brightness of the input state during measurement $m$.  This is used in the reconstruction to account for drifts in the brightness of the source during long acquisitions.  

We consider also the effect of background photons: while the probability of detecting two background photons within the coincidence window $\Delta\tau$ is negligible, coincidences between one signal and one background photon are relevant, given the high fluxes of squeezed and coherent photons. 
We record the background photon rate $b_{i,m}$ at the $i$-th detector and with setting $m$ while the OPO pump and the CS beam are off.
 We then estimate the number of accidental counts due to background photons as $\text{acc}_m^{(i,j)} = [\beta_{i,m}\beta_{j,m} - (\beta_{i,m}-b_{i,m})(\beta_{j,m}-b_{j,m})]\Delta\tau$, where $i,j$ are the indices of the detectors involved in the $m$-th measurement, i.e $\text{acc}_m = \text{acc}_m^{(1,2)}$ for $m=1,4,5,8,9$, $\text{acc}_m = \text{acc}_m^{(3,4)}$ for $m=2$ and $\text{acc}_m = \sum_{i=1}^2\sum_{j=3}^4 \text{acc}_m^{(i,j)}$ for $m=3,6,7,10$.
    
We then use a maximum likelihood estimation (MLE) that takes into account the experimental imperfections described above. Assuming Poisson statistics, we minimize the {weighted error function} $\sum_{m = 1}^{10} [n_\text{th,m}-n_\text{exp,m}]^2/(2n_\text{th,m})$, where the expected coincidence count is calculated as $n_\text{th,m} = \text{Tr}[\Pi_m \rho_{CB}] \eta_m \alpha_m + \text{acc}_m$, where $\eta_m$ {is the efficiency factor describing} the $m$-th outcome, e.g. HH for $m=1$. Considering that when two photons go to a single SPAD no coincidence is recorded, we find $\eta_{1,4,5,8,9} = 2 \gamma_1 \gamma_2$, $\eta_{2} = 2 \gamma_3 \gamma_4$ and $\eta_{3,6,7,10} = (\gamma_1+\gamma_2)(\gamma_3+\gamma_4)$. }

\section*{Statistics}
We used a bootstrapping technique to estimate the error on the density matrices derived with this method. We first generate 100  {10-element lists}, whose $i$-th elements are chosen randomly from a Poissonian distribution with mean $n_\text{exp}[i]$. Then we apply our MLE algorithm to these lists, obtaining 100 density matrices. The distribution of the concurrence shows the effects of statistical errors on the entanglement, showing that even if the number of collected photons was relatively low (on the order of   hundreds of counts), it is sufficient to demonstrate that the concurrence is different from 0 with some standard deviations. The error bars shown in the graphs are centred around the average of the distribution of concurrence and they are {$\pm1$} standard deviations long.

}

~ \\ ~ \\

~ \\ 
\noindent 
~ \\ ~ \\
\noindent 

~ \\ 

~ \\ 

~ \\ 

%


\end{document}